\newif\ifjournal
\journaltrue

\ifjournal
\documentclass{article}
\usepackage[a4paper,total={160mm,240mm}]{geometry}
\usepackage[utf8]{inputenc}
\fi
\usepackage{amsmath}
\usepackage{array,multirow}
\usepackage{hyperref}
\usepackage{nameref}
\usepackage{verbatim}
\usepackage{tikz}
\usepackage{enumitem}
\usepackage{breqn}
\usetikzlibrary{shapes.geometric, arrows,fit, positioning}
\tikzstyle{start} = [rectangle, rounded corners, minimum width=3cm, minimum height=1cm,text centered, draw=black, fill=red!30]

\newcounter{mylabelcounter}

\makeatletter
\newcommand{\labelText}[2]{%
#1\refstepcounter{mylabelcounter}%
\immediate\write\@auxout{%
  \string\newlabel{#2}{{1}{\thepage}{{\unexpanded{#1}}}{mylabelcounter.\number\value{mylabelcounter}}{}}%
}%
}
\makeatother

\title{Cost Optimization of Water Distribution Networks: Model Refinement Is Better Than Problem-Specific Solving Techniques}
\ifjournal
\author{Saumya Goyal, Om Damani, Ashutosh Mahajan}
\fi

\date{November 2021}

\begin{document}

\ifjournal
\maketitle
\fi

\begin{abstract}
Existing techniques for the cost optimization of water distribution networks either employ meta-heuristics, or try to develop problem-specific optimization techniques. %
Instead, we exploit recent advances in generic NLP solvers and explore a rich set of model refinement techniques. 
The networks that we study contain a single source and multiple demand nodes with residual pressure constraints. Indeterminism of flow values and flow direction in the network leads to non-linearity in these constraints making the optimization problem non-convex.
While the physical network is cyclic, flow through the network is necessarily acyclic and thus enforces an acyclic orientation. We devise different strategies of finding acyclic orientations and 
explore the benefit of enforcing such orientations explicitly as a constraint. Finally, we propose a parallel link formulation that models flow in each link as two separate flows with opposing directions. This allows us to tackle numerical difficulties in optimization when flow in a link is near zero. We find that all our proposed formulations give results at par with least cost solutions obtained in the literature on benchmark networks. We also introduce a suite of large test networks since existing benchmark networks are small in size, and find that the parallel link approach outperforms all other approaches on these bigger networks, resulting in a more tractable technique of cost optimization.
\end{abstract}

\ifjournal
\fi

\section{Introduction}\label{intro}

Cost optimization of Water Distribution Networks (WDNs) has been studied for more than six decades~\cite{jetmarova_review}. 
In its most common form, the problem involves finding the least cost way of constructing the network using a given set of pipes while ensuring that pressure and rate of inflow at each demand node meets desired levels. Some variations of the problem consider minimisation of other costs such as those for development, reinforcement and extension. %
Other variants make simplifying %
assumptions about the problem like allowing each link in the network to consist of only a single pipe segment. In \cite{branch}, authors have focused on the cost optimization for acyclic (branched) WDNs that are predominant in rural settings in developing countries, where the luxury of redundancy is often unaffordable. In this work, our focus is on the piping cost optimization of cyclic WDNs.

This problem was initially investigated using techniques of Linear Programming (LP)\cite{twoloop}, Nonlinear Programming (NLP)\cite{su_nlp} and Dynamic Programming\cite{yakowitz_dp}. This was followed by a lot of interest in optimization using metaheuristic techniques
such as Genetic Algorithms~\cite{savic_ga}, Simulated Annealing~\cite{cunha_simulated_annealing}, Ant Colony optimization~\cite{Zheng2017}, 
\ifjournal
Shuffled Frog Leaping\cite{eusuff_shuffled_frog_leaping} and Shuffled Complex Evolution\cite{Liong_shuffled_comlex_evolution}.
\fi

Similar to how advancements in compilers have obviated %
writing assembly code by hand, we believe that currently available advanced optimization solvers obviate the need for development of separate optimization techniques for each problem. 
Our belief is that focus on the development of clever formulations for the problem at hand should give better and faster results when solved on modern solvers. In this work, we propose three equivalent formulations for the piping cost minimisation of cyclic networks which are usually found in the urban setting. %

\begin{figure}[h]
    \centering
    \ifjournal
    \includegraphics[width=0.6\textwidth]{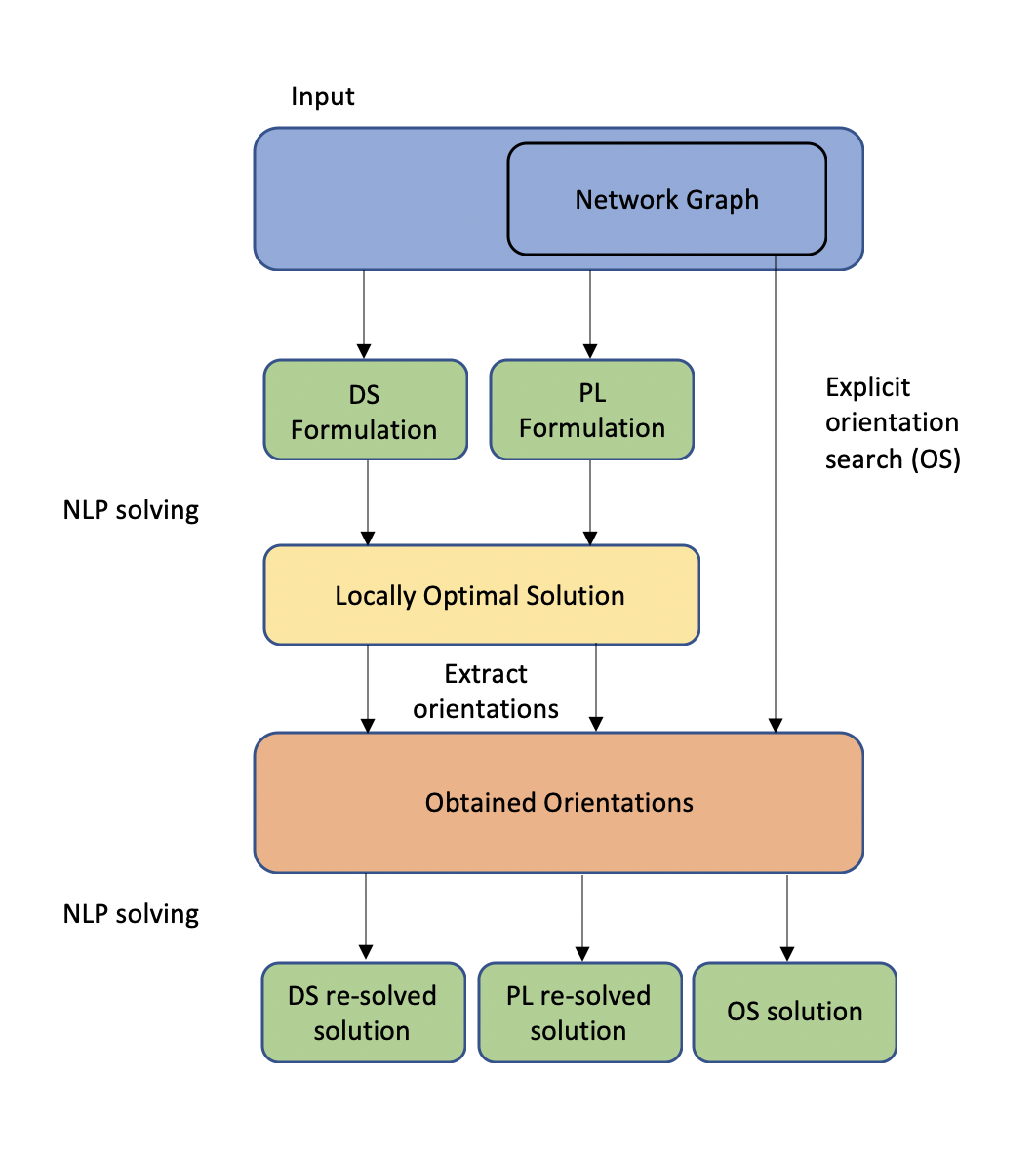}
    \fi
    \caption{
    \ifjournal
    Relationship between techniques developed in this paper.Discrete Segment (DS) and Parallel Link (PL) formulations are constructed from the entire input and solved to give us locally optimal solutions. We can then extract orientations from these solutions to prepare for a re-solving stage. Alternately, we can obtain orientations directly through an explicit search on the network graph giving us a solution from these randomly chosen orientations (Orientation Search, OS)
    \fi}
    \label{fig:formulations}
\end{figure}

We first develop the Discrete Segment formulation that models network constraints without employing any theoretical or experimental insights into the problem. Thanks to recent %
advancements in solvers, optimization on this formulation yields results at par with the least cost solutions obtained by other researchers.
For further improvement, we note that though the physical network is cyclic, water flow through the network is necessarily acyclic.
Finally, we propose the Parallel Link formulation that overcomes numerical difficulties in optimization when flow in a link is near zero by modelling flow through links in opposing directions using different flow variables. We evaluate these methods on popular benchmark networks. Since these benchmark networks are simple in nature, we also introduce a new test suite consisting of 6 representative networks from the HydroGen~\cite{hydrogen} archive for large scale comparison of the proposed formulations. We find that Parallel Link outperforms all other formulations
and techniques
for cyclic water network cost optimization. Fig \ref{fig:formulations} provides a summary of the discussed formulations and the solving process employed for each of them. 

\ifjournal
We have incorporated our proposed formulation into  JalTantra~\cite{hooda2019}, a system for WDN optimization.
\fi
\section{Problem Formalisation}
\ifjournal
As noted in Section \ref{intro}, we focus on piping cost optimization of single source Water Distribution Networks. These WDNs can be modeled using network flow graphs.
\fi
The vertices in these graphs represent 
\ifjournal
either source or demand nodes.
\fi
The demand nodes 
\ifjournal
in the network 
\fi
are specified along with the their demand rates, node elevation and residual pressure constraints. Residual pressure constraint for each node reflects the least pressure water should have at the node, above the node elevation.

The edges in the graph represent links along which pipes are to be
\ifjournal
laid to construct the network.
\fi
Since real networks deploy multiple segments of varying diameters along a single link, our formulations also allow a single link to be constructed with multiple pipe segments. This is called the ``Split Pipe" approach. Many works use what is called the ``Discrete Pipe" approach which allow each link to be constructed with only a single pipe and hence, in general, lead to more restrictive, less practical and higher cost solutions. The pipes that we can use for construction of the network are given as the set of commercially available pipes which differ in their diameters, cost per unit length and roughness coefficients. Pipes of smaller diameters are cheaper but offer more resistance to the flow of water. 

The notation followed through the paper is as follows.

\subsection{Inputs} \label{inputs}
\begin{itemize}
    \item $G(N,L)$: 
    \ifjournal
    The network graph, undirected and cyclic. Here $N$ is the set of nodes in the graph and $L$ is the set of links.
    \fi
    \item $s \in N$: Unique source node
    \item $E_i$ for $i \in N$: Node elevation %
    \ifjournal
    \item $D_i$ for $i \in N/\{s\}$: Node demand
    \item $p_i$ for $i \in N/\{s\}$: Additional minimum pressure required at node. The minimum pressure required at each node is then expressed as $p_i + E_i$
    \fi
    \item $l_i$ for $i\in L$: Link length
    \item $P$: Set of all commercially available pipes %
    \item $q_m$ and $q_M$: Minimum and maximum value of absolute flow allowed through each
    \ifjournal
    link, assumed independent of link position and pipe diameter
    \fi
\end{itemize}{}

\subsection{Outputs}\label{outputs}
$l_{ij}$ for $i \in L, j \in P$ representing length of $j^{th}$ commercially available pipe type (henceforth referred to as pipe type) segment in link $i$ 
\ifjournal
in the least cost network obtained %
\fi

\subsection{Objective Function} \label{capital_cost}
Capital cost incurred for network construction is equal to the sum of costs associated with each pipe segment in each link:
\begin{equation*}
    C(.)=\sum_{i \in L}\sum_{j \in P} l_{ij}*c_j
\end{equation*}

\subsection{Constraints} \label{constraints}
\begin{itemize}
    \item \textbf{Residual pressure: }water reaching demand nodes should have pressure higher than the minimum pressure requirement at these nodes. Thus if $h_i$ is the head loss from root to node $i$, then $h_i<E_s-E_i-p_i, \forall i \in N$. The loss in pressure due to friction (headloss) across a pipe of diameter $d$, length $l$ and roughness coefficient $R$ is calculated using the Hazen-Williams equation\cite{williams1933} :
    \begin{equation*}
        h=\frac{\omega * l * q^{1.852}}{R^{1.852} * d^{4.87}}
    \end{equation*}
    \ifjournal
    The constant $\omega=10.68$ when $q$ is expressed in units of $m^3/s$ and $d$, $l$ and $h$ are in units of \textit{metres}.
    \fi
    It is important to note that $q$ above must be non-negative and there is a loss in pressure towards the direction of flow.
    \item \textbf{Flow range: }absolute value of flow 
    \ifjournal
    in any link 
    \fi
    should be between the maximum ($q_M$) and the minimum ($q_m$) tolerable values. Thus if $q_i$ is the flow in $i^{th}$ link then $q_m \leq |q_i| \leq q_M, \forall i \in L$.
\end{itemize}
\ifjournal
Given the inputs to the problem defined in Section \ref{inputs} the overall objective is to give values for the variables defined in Section \ref{outputs} to obtain the least construction cost of the WDN as defined in Section \ref{capital_cost} while satisfying constraints as defined in Section \ref{constraints}.
\fi
\section{Solution Approach}
Our objective is to develop optimization formulations to be optimised directly using off-the-shelf solvers. We could devise a raw formulation directly from the problem statement, but it is challenging to devise one that can give near-optimal solutions in a time constrained setting. %
If care is not taken, the many different constraints being applied on the many different paths in the network graph can easily overwhelm the solver with their sheer number.

Many different formulation techniques employing theoretical results on WDNs have been devised by many authors. In \cite{two_pipes} it is proved that each link can be constructed with at most 2 segments from pipes of adjacent diameters in an optimal solution, provided cost per unit length is a convex function of the diameter, which is readily met in real life settings. Using this result, a \textbf{Continuous Segment} formulation is proposed in \cite{cont_seg} in which each link is constructed with a single pipe whose diameter can take any value from the least to the highest diameter. After solving, the pipe diameter obtained for each link can be broken into two adjacent commercially available pipe diameters which offer the same headloss as the original pipe. Approximating the cost per unit length and the roughness coefficient as a function of the diameter poses a challenge in this approach which often leads to non-optimality. 

Another alternative is to directly impose the 2 adjacent pipe result as a constraint to the solver resulting in an MINLP formulation which is generally harder to optimise than an NLP formulation. In addition, in \cite{branch} it was found that incorporating the two adjacent pipe constraints for branched (acyclic) input networks results in a much longer running time. While our focus is on cyclic networks, the factors for performance degradation are applicable in our setting as well and hence we do not explore this further.

We now discuss three main ideas for converting the problem into non-linear programming problems. In contrast to the techniques discussed above, we do not employ any previously known theoretical results in our formulations.

\subsection{Discrete Segment Formulation} \label{ds}
This formulation is based on a model that reduces the number of constraints passed to the solver by reducing the number of paths considered while calculating headloss. The name serves as a contrast to alternative ways of formulating the problem which were discussed above.
\ifjournal

\fi
Below is a description of the Discrete Segment formulation in the mathematical form it is passed to the solver.
\subsubsection{Formulation}
\begin{itemize}
    \item \textbf{Optimization Variables}
    \begin{itemize}[leftmargin=0em]
        \item \textbf{Flow variables:} 
        Variables $q_i$ for $i \in L$ denote flow through each link in the network. Since water can flow in either direction in each link, we need to define the direction of a "positive" flow \ifjournal
        ie, the flow direction for which the corresponding flow variable is positive
        \fi
        . This choice of directions is described later in \nameref{const:flow_cons}. Flow variables are bounded by the \textbf{Flow range} constraint as: $q_m \leq |q_i| \leq q_M, \forall i \in L$.
    \end{itemize}
    
    \begin{itemize}[leftmargin=0em]
        \item \textbf{Segment length variables:} We formulate each link to be constructed with segments of all the pipes types with discrete diameter values. The variable $l_{ij}$ denotes the length of the $j^{th}$ pipe segment in the $i^{th}$ link in the graph as described in Section \ref{outputs}. Segment length variables are bounded as: $0 \leq l_{ij} \leq l_i, \forall i \in L, \forall j \in P$. The \textbf{Sum of link segments} constraint ensures that sum of all segments of a link are equal to the total length of the link: $\sum_{j\in P} l_{ij} = l_i, \forall i \in L$.
    \end{itemize}

    \item \textbf{Cost Function}
    
    The objective cost for optimization is set to be equal to the capital cost for network construction.
    
    \begin{equation*}
        C(.)=\sum_{i\in L} \sum_{j \in P} l_{ij} * c_j
    \end{equation*}{}
    
    \item \textbf{Constraints}
    
    \begin{itemize}[leftmargin=0em]

        \item \textbf{\labelText{Flow conservation}{const:flow_cons}.}The difference of the inflow and outflow to any node should be equal to its demand. To formulate this constraint, we introduce the matrix $\mathcal{F}$, indexed by $N \times L$ that disambiguates the meaning of positive flow in each link.
        
        We arbitrarily define the direction of positive flow for each link in the network by assigning one of the endpoints as the sink node for the link. The $(i,j)^{th}$ entry of the matrix ($\mathcal{F}_{i,j}$) is thus set to 1 or -1 depending on whether or not node $i$ is the sink for link $j$. It is 0 if link $j$ is not connected to node $i$.
        
        The flow conservation constraint is formulated as: 
        \begin{equation*}
            \sum_{j\in L} \mathcal{F}_{i,j}*q_j = D_i, \forall i \in N
        \end{equation*}{}
        
        \item \textbf{Residual Pressure.} This constraint described in Section \ref{constraints} requires that water reaching demand nodes should have pressure higher than the minimum pressure requirement. Since we have modeled flow through each link independently, pressure at any node has to be calculated using headloss equations for paths beginning at source nodes (having known pressure), ending at demand nodes (having unknown pressure). We break this constraint into the following two constraints which reduces the number of paths considered.
        
        \item \textbf{\textit{Pressure drop across a cycle.}} Each node should have a unique value of pressure and hence the pressure drop along every cycle in the graph must be 0. However, there can be an exponential number of cycles
        \ifjournal
        in a given graph in the worst case 
        \fi
        and hence an exponential number of constraints enforced on the solver.

        Interestingly, we do not need to enforce the 0 pressure drop constraint on every cycle in the graph but only on any \textbf{cycle basis} of the graph. 
        A cycle basis is a set of cycles in the graph such that any cycle in the graph can be formed by a summation over some subset of the cycle basis. Here a summation of cycles in a set results in a subgraph containing 
        \ifjournal
        exactly those 
        \fi
        edges which occur in exactly odd number of cycles in the set.
        
        A cycle basis can be constructed by considering the cycles formed by all the $(|L|-|N|+1)$ edges which are not present in the spanning tree, with the spanning tree\cite{cycle_bases}. The set of cycles thus formed is denoted by $C$.
        
        We formulate this constraint using Hazen-Williams eqn as:
        \begin{equation*}
            \sum_{j\in L} \sum_{k\in P} \frac{\mathcal{C}_{i,j} * \omega * l_{jk}*q_j*|q_j|^{0.852}}{R_k^{1.852} * d_k^{4.87}} = 0, \forall i \in C
        \end{equation*}{}
        
        $\mathcal{C}$ is a matrix indexed by $C \times L$. Note that $|C|=|L|-|N|+1$. 
        The matrix has entries from 1,0,-1. $(i,j)^{th}$ entry of the matrix ($\mathcal{C}_{i,j}$) is 1 if the direction of walk of link $j$ in cycle $i$ is the same as the direction assigned to link $j$ %
        , -1 if it is in the opposite direction and 0 if link $j$ is not present in cycle $i$.
        \item \textbf{\textit{Head at node.}} The previous constraint ensures null pressure drop across all cycles thus guaranteeing independence of pressure drop across two nodes from the walk taken between the nodes. This allows us to specify constraints for the pressure at a node using only 1 path from the root to the node, which is chosen as the shortest path along a spanning tree of the graph.
        This constraint is formulated using Hazen-Williams eqn as: 
        \begin{displaymath}
            0\leq \sum_{j\in L} \sum_{k\in P} \frac{\mathcal{S}_{i,j} * \omega * l_{jk}*q_j*|q_j|^{0.852}}{R_k^{1.852} * D_k^{4.87}}  \leq E_s - E_i-P_i, \forall i\in N
        \end{displaymath}{}
        $\mathcal{S}$ is a matrix indexed by $N\times L$ with entries from 1,-1,0. We construct a spanning tree of the graph and consider paths from source to each node in the spanning tree. $(i,j)^{th}$ entry of the matrix ($\mathcal{S}_{i,j}$) is 1 or -1 depending on whether the arbitrary direction assigned to link $j$ is towards node $i$ in the path under consideration %
        . It is 0 if link $j$ is not in the path.

    \end{itemize}
\end{itemize}

We will see in Section \ref{results} that this formulation is at par with the best results known in the literature. We now look at potential sources of improvement on the Discrete Segment formulation.

\subsection{Breaking into orientations}
This formulation separates the solving process for flow value and flow direction by modelling them as separate problems. The motivation for this comes from the fact that fixing flow directions reduces the range of values flow variables can take and thus results in a reduced system which is easier to solve and allows us to obtain better solutions from the solver. Overall, 
we see that the only difference of this approach with discrete segment is that both problems are being solved explicitly here.

\ifjournal
We begin by noting that the flow in a water distribution network is acyclic which follows from Pressure drop across cycles constraints. 
\fi
In \cite{orientations} it is claimed that expressing this explicitly to a non-convex global optimality solver makes the solving process for gas networks simpler. We exploit this observation in the context of water networks.

An orientation is defined as a map that assigns a direction to flow in each edge in the graph. Note that fixing an orientation still doesn't determine the exact flow value through each link in the graph so that a solution for the problem would correspond to flow values through each link and lengths of pipe segments satisfying all constraints, in addition to the orientation.

There exist several techniques 
\ifjournal
(e.g. \cite{enumeration})
\fi
for enumeration of all possible orientations of a graph. The techniques for gas networks are based on following a reduction process of the graph centered at collapsing paths with zero-demand nodes. But, we found that such nodes %
are rarely present outside of degree 2 paths in water networks.

We now discuss methods to tackle both parts of the problem for water distribution networks: finding orientations and finding solutions given an orientation. 

\subsubsection{Finding Orientations Algorithmically}\label{sec:find_or}
One way to find orientations is to perform an algorithmic search on the network graph itself. A naive way to do this would be to consider all permutations of nodes since each permutation of the nodes represents a topological sort and so corresponds to an acyclic orientation. However:
\begin{enumerate}
    \item Different permutations can correspond to same orientations%
    \item Many permutations will lead to some non-source nodes having no incoming links, which is invalid
\end{enumerate}
\ifjournal
\begin{figure}[h]
    \centering
    \includegraphics[width=\textwidth]{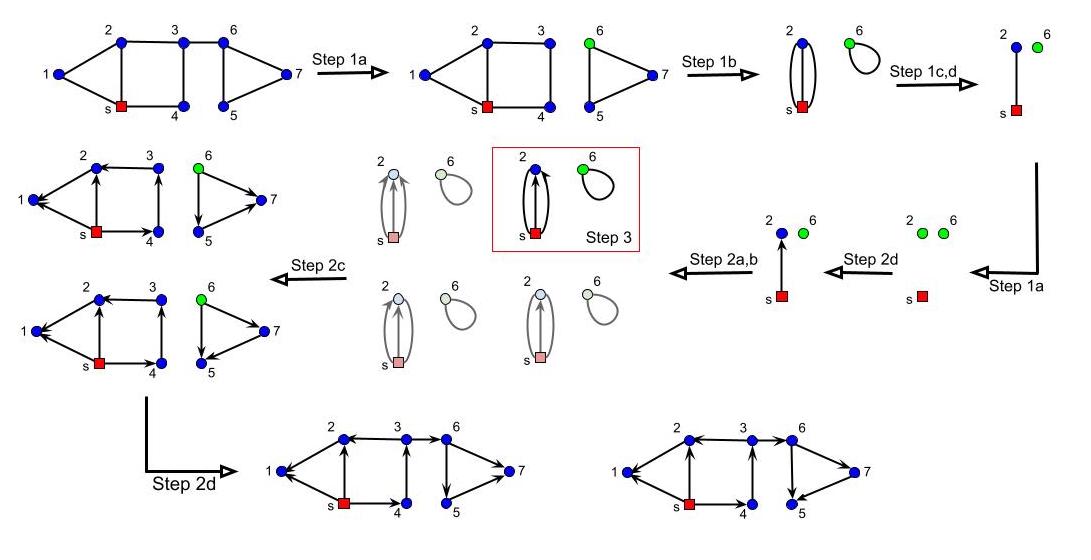}
    \caption{Orientation Search on a sample graph}
    \label{fig:graph_reduction}
\end{figure}
\fi
From these two facts it is clear that we cannot exploit permutations to generate orientations directly. Moreover, as noted, we do not find many zero-demand nodes outside of degree 2 paths in water networks. We thus devise a graph reduction process centered at removing parallel edges and degree 2 nodes to obtain orientations efficiently. 
\ifjournal
\fi
\ifjournal

The algorithm is described below and illustrated in Figure \ref{fig:graph_reduction}. The red node in the figure represents the source and green nodes represent sources for connected components. Steps that could not be applied at specific stages have been omitted from the figure for brevity.

\begin{enumerate}
    \item \textbf{Reduction.} Graph reduction is performed in a recursive manner, reducing the number of edges in the graph with each pass. Each connected component in the resulting simple graphs has a defined source for that component. The following steps are repeated till they do not result in a reduction of the number of edges in the graph.
    
    \begin{enumerate}
        \item Remove all bridges from the graph since flow directions are fixed in such links. Removal of a bridge increases the number of connected components by 1, and the cut-vertex corresponding to the removed bridge in the latest connected component is made the source node for that connected component.
        \item Remove all degree 2 nodes and collapse paths corresponding to the removed nodes into single edges between the end points. (Note that, it is possible that at some point during reduction we end up with a connected component containing only a single cycle. This cycle is collapsed into a self-loop at the source.)
        \item Remove all self-loops.
        \item Collapse all parallel edges into single edges resulting in a simple graph.
    \end{enumerate}
    Since water networks typically have low connectivity, we expect the final reduced graph to be small enough to allow enumeration of all feasible orientations. %
    Thus, we exhaustively generate all orientations of all connected components of the final reduced graph and discard infeasible orientations (i.e., either it is a  cyclic orientation or, the orientation has non-source nodes with no incoming links or, the source node has incoming links in the orientation).

    \item \textbf{Expansion.} Following the reduction process, each orientation of the reduced graph is expanded recursively, in the reverse order of reduction.
    \begin{enumerate}
        \item Parallel edges are re-introduced and each is assigned flow either in the same direction as the collapsed edge or given no direction at all. Giving an edge no direction during expansion is allowed when the edge will eventually be expanded into a path of degree 2 nodes with one node as a sink.
        \item Re-introduce self-loops. Since self-loops expand into cycles, they are given no direction when they are re-introduced.
        \item Re-introduce the removed degree-2 nodes and expand the collapsed paths. If the collapsed edge had no direction assigned to it, then one of the  intermediate nodes is made a sink. Otherwise all edges in the path are given flow in the same direction as the collapsed edge.
        \item Re-introduce bridges with appropriate directions assigned to them.
    \end{enumerate}
    
    \item \textbf{Selecting orientations.} This reduction followed by expansion process allows us to enumerate valid orientations of the graph. It was expected that the low connectivity of the input graph would result in only a small number of valid orientations. Though this generally held true in the small networks commonly used in literature, a large number of orientations indeed turned out to be feasible for the new synthetic networks we experimented with. Due to this, instead of enumerating all valid orientations, we choose to randomly select a subset of the orientations. Simply sampling a subset of the possibilities during each phase instead of expanding all possible orientations allows the chosen orientations to be sufficiently different from each other, giving a representative subset of all orientations. %
    We call this method \textbf{Orientation Search (OS)}.
\end{enumerate}

\fi
\subsubsection{Finding Orientations from Existing Solutions}
In addition to finding orientations through an algorithmic approach on the input graph, we also experiment with using the orientations obtained from solutions of the other suggested optimization formulations. By finding solutions to 
\ifjournal
the orientations thus obtained, we are effectively \textbf{Re-Solving (RS)} these orientations.
\fi

\subsubsection{Finding Solutions with Given Orientations} \label{sssec:sol_to_or}
Solving for optimality given an orientation is a relatively simpler process. An orientation effectively forces a sign on the flow variables for each link and hence allows us to use the same cost function and constraints as in the previous section.

We see in Section \ref{results} that
\ifjournal
a random selection of a subset of orientations
\fi
doesn't work well since too many orientations are infeasible and result in no solution. The Re-Solving approach however does give better results than Discrete Segment and can thus be used to refine output obtained through Discrete Segment.

\subsection{Parallel Link Formulation} 
In the previous section we discussed a method of separating the solution process for flow value from flow direction by fixing the direction for flow through orientations. We now discuss a method of separating this solution process by modelling flow through a pipe using two separate flow variables. These variables will be denoted as $q_{i1}$ and $q_{i2}$ for each link $i$ and are constrained to have non-negative values such that $q_{i1}$ is positive when $q_i$ would be positive and $q_{i2}$ is positive when $q_i$ would be negative. Effectively it can be imagined that each link is replaced by two links which only allow flow in either direction.

Note that the notion of introducing two parallel link in place of every link is only to aid in understanding of how the introduced variables behave. The resulting variables $l_{ij}$ are interpreted 
\ifjournal
just as they were interpreted in the above formulations to construct the network 
\fi
and the resulting network does not actually comprise of such parallel link.

\ifjournal
\subsubsection{Formulation}
The only changes in formulation from Discrete Segment are:
\begin{itemize}

    \item \textbf{Variable bounds:} $q_m \leq q_{i1}, q_{i2} \leq q_M, \forall i \in L$

    \item \textbf{Constraints}
    \begin{itemize}[leftmargin=0em]
        \item \textbf{\labelText{Unique flow direction}{cons:unique_flow_dir_j}.} Since $q_i$ could either have been positive or negative or 0, at most one of $q_{i1}$ and $q_{i2}$ should ever be non-zero.
        \begin{equation*}
            q_{i1}*q_{i2} = 0, \forall i \in L
        \end{equation*}{}

        Since at most one of the two variables will ever be non-zero, we can replace $q_i$ with the difference of $q_{i1}$ and $q_{i2}$ in other constraints. They are formulated as given below
        
        \item \textbf{Flow conservation.}
        \begin{equation*}
            \sum_{j\in L} \mathcal{F}_{i,j}*(q_{j1}-q_{j2}) = D_i, \forall i \in N
        \end{equation*}{}
        
        \item \textbf{Headloss at node.}
        \begin{equation*}
            0\leq \sum_{j\in L} \sum_{k\in P} \frac{\mathcal{S}_{i,j} * \omega * l_{jk}*(q_{j1}^{1.852}-q_{j2}^{1.852})}{R_k^{1.852} * d_k^{4.87}} \leq E_s - E_i-P_i, \forall i\in N
        \end{equation*}{}
        
        \item \textbf{Pressure drop along cycle.}
        \begin{equation*}
            \sum_{j\in L} \sum_{k\in P} \frac{\mathcal{C}_{i,j} * \omega * l_{jk} * (q_{j1}^{1.852} - q_{j2}^{1.852})} {R_k^{1.852} * d_k^{4.87}} = 0, \forall i \in C
        \end{equation*}{}
    \end{itemize}
\end{itemize}
\fi
\ifjournal
We see in Section \ref{results} that this formulation performs 
\fi
the best among all proposed formulations which can be attributed to its handling of numerical difficulties for optimization when flow is near zero. %

\section{Experiments and Results}\label{results}
\subsection{Dataset}
The aforementioned formulations were implemented using the local optimality solver Ipopt \cite{ipopt} and tested on various parameters of performance. In addition to single-source benchmarking networks Two-loop \cite{twoloop}, Taichung \cite{sung2007}, Hanoi \cite{hanoi}, Double Hanoi and Triple Hanoi \cite{Cisty2010}, we incorporate test networks from the HydroGen \cite{hydrogen} archive to perform a comparison of our formulations on larger inputs. Benchmark comparisons performed in Table \ref{table:comparison} demonstrate that the Discrete Segment approach is at par with the best approaches in literature on the simple, and generally small networks used in literature.

A comparison of re-solving Discrete Segment (RS), Parallel Link (PL) and Orientation Search(OS) can be found in Table \ref{table:resolve_pl_ro}.
\ifjournal
We note key observations from these and the other techniques discussed in the paper in the sections below.
\fi
\ifjournal
\begin{table}[]
\fi
\centering
\makebox[\textwidth]
{
    \begin{tabular}{|c|c|c|c|c|c|c|c|c|}
    \hline
    Method & Year & TwoLoop & Hanoi %
    & D.Hanoi & T.Hanoi & Taichung %
    & Solution type\\
    \hline
    DS & - & \textbf{4.04$\times 10^5$} & \textbf{6.06$\times 10^6$} %
    & \textbf{1.21$\times 10^7$} & \textbf{1.84$\times 10^7$} & \textbf{8.77$\times 10^6$} %
    & Split Pipe\\
    \ifjournal
    \hline
    Gen. Disj. MINLP\cite{cassiolato_2021} & 2021 & 4.19$\times 10^5$ & - & - & - & - & Discrete Pipe\\
    \fi
    \hline
    TPS\cite{manolis2021} & 2021 & 4.19$\times 10^5$ & 6.08$\times10^6$ & - & - & - & Discrete Pipe\\
    \hline
    OPUS\cite{saldarriaga_2020} & 2020 & - & 6.37$\times 10^6$ & - & - & 8.91$\times 10^6$ & Discrete Pipe\\
    \ifjournal
    \hline
    ABC\cite{yilmaz2020} & 2020 & 4.29$\times 10^5$ & 6.12$\times10^6$ & - & - & - & Discrete Pipe\\
    \hline
    WOA\cite{ezzeldin2020} & 2020 & 4.19$\times 10^5$ & 6.08$\times10^6$ & - & - & - & Discrete Pipe\\
    \fi
    \hline
    GA+flow-penalty\cite{abbas2019} & 2019 & 4.19$\times 10^5$ & 6.08$\times10^6$ & - & - & - & Discrete Pipe\\
    \ifjournal
    \hline
    WCANET\cite{praneeth2019} & 2019 & 4.19$\times 10^5$ & 6.12$\times10^6$ & - & - & - & Discrete Pipe\\
    \fi
    \hline
    Disj. MINLP\cite{caballera2019} & 2019 & 4.19$\times 10^5$ & 6.08$\times10^6$ & - & - & - & Discrete Pipe\\
    \ifjournal
    \hline
    ACO adaptive\cite{bahoosh2019} & 2019 & - & 6.41$\times10^6$ & - & - & - & Discrete Pipe\\
    \fi
    \hline
    $\text{ACO}_{\text{CTC}}$\cite{Zheng2017} & 2017 & - & 6.08$\times 10^6$ %
    & - & - & -  %
    & Discrete Pipe\\
    \ifjournal
    \hline
    Mod. PSO\cite{surco2017} & 2017 & 4.19$\times 10^5$ & 6.08$\times10^6$ & - & - & - & Discrete Pipe\\
    \hline
    EAA-WDND\cite{Avila-Melgar2016}& 2016 & 4.19$\times 10^5$ & - %
    & - & - & - %
    & Discrete Pipe \\
    \fi
    \hline
    CSHS\cite{Sheikholeslami2015} & 2015 & - & 6.08$\times 10^6$ %
    & 1.23$\times 10^7$ & - & - %
    & Discrete Pipe\\
    \hline
    STA\cite{Zhou2015} & 2015 & 4.19$\times 10^5$ & 6.10$\times 10^6$ %
    & - & \textbf{1.84$\times 10^7$} & -  %
    & Discrete Pipe\\
    \hline
    OPUS+Meta-heur.\cite{Saldarriaga2014} & 2014 & - & 6.08$\times 10^6$ %
    & - & - & 8.81$\times 10^6$  %
    & Discrete Pipe\\
    \ifjournal
    \hline
    IMBA\cite{Sadollah2014} & 2014 & - & 6.08$\times 10^6$ %
    & - & - & - %
    & Discrete Pipe\\
    \hline
    PHSM\cite{Bi2015} & 2015 & - & 6.11$\times 10^6$ %
    & - & - & - %
    & Discrete Pipe\\
    \hline
    ALCO-GA\cite{Johns2013} & 2013 & 4.19$\times 10^5$ & - %
    & - & - & -  %
    & Discrete Pipe\\
    \hline
    PSO-DE\cite{SEDKI2012} & 2012 & 4.19$\times 10^5$ & 6.08$\times 10^6$ %
    & - & - & - %
    & Discrete Pipe\\
    \hline
    MINLP\cite{Bragalli2012} & 2012 & 4.19$\times 10^5$ & 6.11$\times 10^6$ %
    & - & - & - %
    & Discrete Pipe\\
    \fi
    \hline
    GALP\cite{Cisty2010} & 2010 & - & \textbf{6.06$\times 10^6$} %
    & \textbf{1.21$\times 10^7$} & \textbf{1.84$\times 10^7$} & - %
    & Split Pipe\\
    \hline
    GA-LP\cite{Krapivka-2009} & 2009 & \textbf{4.04$\times 10^5$} & - %
    & - & - & - & Split Pipe\\
    \hline
    TS\cite{sung2007} & 2007 & - & 6.08$\times 10^6$ & - & - & \textbf{8.77 $\times 10^6$} & Discrete Pipe\\
    \ifjournal
    \hline
    Flow search+optimization\cite{Loganathan-1995} & 1995 & \textbf{4.04$\times 10^5$} & - %
    & - & - & - & Split Pipe\\
    \fi
    \hline
    \end{tabular}
}
\label{table:comparison}
\ifjournal
\caption{Performance comparison of DS with existing techniques}
\fi
\ifjournal
\end{table}
\fi

\subsection{Orientation Search}\label{sec:os_results}
We note in Section \ref{sec:find_or} that conditions of acyclicity and non-zero inflow to non-source nodes are satisfied by too many orientations in a network to allow enumeration. 
Hence, instead of enumerating all orientations, we randomly sample orientations from different subspaces of the orientation space
\ifjournal
as explained in Section \ref{sec:find_or} 
\fi
to ensure that the selected set is representative of the entire set. 

Observe from Table \ref{table:resolve_pl_ro} that Orientation Search performs a lot better in terms of time than the other approaches on the benchmarking networks. This can be attributed to the reduction of the search space in size exponential to the number of links present in the graph due to the fixing of orientations (the flow direction for each link is fixed and hence the flow variables can only admit either positive or negative values). It performs fairly well in terms of the value obtained as well and hence looks promising for small test cases. The last column in Table \ref{table:resolve_pl_ro} shows the number of feasible orientations found out of the randomly selected 100 orientations (except for TwoLoop, where only 9 orientations are possible). For large graphs, the Orientaion Search approach fails entirely and all orientations found using this approach are actually infeasible implying that feasible orientations are still very sparse within acyclic orientations that have non-zero inflow to non-source nodes. 

This makes the process of generating random orientations from the graph structure infeasible to use for 
\ifjournal
larger and real sized 
\fi
networks. One thus has to rely on other techniques for finding orientations.

\subsection{Re-Solving Discrete Segment Orientations}\label{ssec:rsds}
To look at the performance of re-solving orientations obtained through discrete segment, we observe 100 runs of the discrete segment formulation with randomised starting points and re-solve for each of the orientations obtained through discrete segment. Cost optimization for the re-solving approach is performed as many times for each orientation as the number of times that particular orientation was found in the discrete segment formulation. Essentially, we rely on the solver to not only come up with feasible orientations that admit better solutions, but also to come up with better orientations more often. We hence spend more computational effort on re-solving better orientations to get better results.

We 
observe that although there is no improvement that re-solving offers in terms of least cost obtained through the orientations, the standard deviation and the average cost obtained through re-solving are usually smaller. It is noteworthy that these improvements over the discrete segment formulation are subject to having already run the discrete segment formulation to obtain the orientations in the first place. Improvements in average and standard deviation cost and time taken warrant using a lesser number of runs of the discrete segment which are subject to re-solving than to perform a higher number of runs of discrete segment which are not subject to re-solving, in a time constrained setting.

\ifjournal
\begin{table}
\centering
\fi
\makebox[\textwidth]
{
\begin{tabular}{|c|c|c|c|c|c|c|c|c|c|c|c|c|c|c|c|}
\hline
    \multicolumn{3}{|c|}{Test Case} & \multicolumn{3}{c|}{min ($10^6$)} & \multicolumn{3}{c|}{avg ($10^6$)} & \multicolumn{3}{c|}{std ($10^5$)} & 
    \multicolumn{3}{c|}{total time (1000s)} & \multirow{2}{4em}{OS-\#feasible}\\
    \cline{1-15}
    Name & Nodes & Links & RS & PL & OS & RS & PL & OS & RS & PL & OS & RS & PL & OS & \\
    \hline
    TwoLoop & 7 & 8 & \textbf{0.40} & \textbf{0.40} & \textbf{0.40} & \textbf{0.46} & 0.55 & 0.48 & \textbf{0.77} & 1.38 & \textbf{0.77} & %
    0.18 & 0.16 & \textbf{0.006} & 9\\
    \hline
    Taichung & 20 & 31 & \textbf{8.76} & 8.77 & 9.40 & \textbf{9.14} & 9.26 & 10.3 & \textbf{1.92} & 3.59 & 5.20 & %
    2.44 & 2.21 & \textbf{1.17} & 100 \\
    \hline
    Hanoi & 32 & 34 & \textbf{6.06} & \textbf{6.06} & 6.07 & 6.22 & \textbf{6.18} & 6.63 & 0.96 & \textbf{0.95} & 4.80 & %
    1.39 & 1.21 & \textbf{0.7} & 44\\
    \hline
    D.Hanoi & 62 & 67 & \textbf{12.1} & \textbf{12.1} & 12.6 & 12.4 & \textbf{12.3} & 13.3 & 1.68 & \textbf{1.46} & 6.31 & %
    9.53 & 7.63 & \textbf{3.59} & 25 \\
    \hline
    T.Hanoi & 92 & 100 & \textbf{18.4} & \textbf{18.4} & 18.9 & 18.9 & \textbf{18.7} & 20.4 & \textbf{1.74} & 1.76 & 10.2 & %
    26.9 & 23.1 & \textbf{12.4} & 13 \\
    \hline
    SP\_1\_4 & 73 & 100 & 3.56 & \textbf{3.55} & - & 3.70 & \textbf{3.66} & - & 1.2 & \textbf{1.0} & - & %
    55.1 & \textbf{45.9} & - & 0\\
    \hline
    SP\_2\_3 & 78 & 100 & \textbf{3.40} & \textbf{3.40} & - & 3.53 & \textbf{3.50} & - & 2.2 & \textbf{2.0} & - & %
    67.7 & \textbf{33.4} & - & 0\\
    \hline
    SP\_3\_4 & 83 & 99 & \textbf{6.92} & \textbf{6.92} & - & 6.99 & \textbf{6.97} & - & 0.47 & \textbf{0.37} & - & %
    83.9 & \textbf{60.0} & - & 0\\
    \hline
    SP\_4\_2 & 143 & 198 & 9.01 & \textbf{8.84} & - & 9.26 & \textbf{9.15} & - & 1.78 & \textbf{1.62} & - & %
    75.3 & \textbf{46.9} & - & 0\\
    \hline
    SP\_5\_5 & 155 & 200 & 9.42 & \textbf{9.37} & - & \textbf{9.71} & 9.87 & - & \textbf{1.89} & 3.98 & - & %
    806 & \textbf{387} & - & 0\\
    \hline
    SP\_6\_3 & 166 & 198 & 10.9 & \textbf{10.7} & - & 11.5 & \textbf{11.0} & - & 4.92 & \textbf{2.98} & - & %
    743 & \textbf{338} & - & 0\\
    \hline
     
\end{tabular}%
}
\label{table:resolve_pl_ro}
\ifjournal
\caption{Comparison of minimum, average and standard deviation of costs obtained in 100 runs using Re-solving of discrete segment (RS), Parallel Link (PL) and Orientation Search (OS) formulation.}
\fi
\ifjournal
\end{table}
\fi

\subsection{Parallel Link}\label{sec:pl_results}
The parallel link approach is tested by starting all runs with the same starting point as for discrete segment. We observe that Parallel Link results in smaller minimum cost values for bigger test cases and smaller standard deviation and average cost than re-solving on most test networks. Time taken by the parallel link formulation is also 
\ifjournal
comparable to that by the discrete segment formulation (and hence 
\fi
smaller than overall running time of discrete segment + re-solving on discrete segment
\ifjournal
)
\fi
. This makes Parallel Link a clearly better choice. As we observe in Table \ref{table:comparison} that Discrete Segment is at par with the literature, we can conclude that parallel link is the best known approach for cost optimization of WDNs.

\ifjournal
\begin{table}
\begin{center}
\begin{tabular}{|c|c|c|c|c|c|c|}
\hline
    \multirow{2}{*}{Test Case} & \multicolumn{2}{c|}{\# succ. runs} & \multicolumn{2}{c|}{\# dist. or.} & \multicolumn{2}{c|}{\# comm. links} \\
    \cline{2-7}
     & DS & PL & DS & PL & DS & PL\\
    \hline
    TwoLoop & 94 & 100 & 9 & 7 & 3 & 3\\
    \hline
    Taichung & 98 & 100 & 95 & 98 & 2 & 4\\
    \hline
    Hanoi & 100 & 100 & 22 & 19 & 23 & 24\\
    \hline
    D.Hanoi & 97 & 97 & 62 & 64 & 44 & 48\\
    \hline
    T.Hanoi & 98 & 98 & 90 & 86 & 66 & 72\\
    \hline
    SP\_1\_4 & 96 & 94 & 22 & 19 & 81 & 90\\
    \hline
    SP\_2\_3 & 98 & 99 & 24 & 16 & 87 & 90\\
    \hline
    SP\_3\_4 & 97 & 96 & 4 & 1 & 96 & 99\\
    \hline
    SP\_4\_2 & 89 & 100 & 83 & 83 & 149 & 151\\
    \hline
    SP\_5\_5 & 84 & 99 & 72 & 74 & 156 & 168\\
    \hline
    SP\_6\_3 & 87 & 96 & 48 & 36 & 180 & 182\\
    \hline
\end{tabular}
\caption{Comparison of solutions obtained through Discrete Segment and Parallel Link on different testcases}
\label{table:details}
\end{center}
\end{table}
\fi

\ifjournal
In order to investigate the causes for this improvement, we enumerate the following quantities related to the output of 100 random runs of discrete segment and parallel link formulations in Table \ref{table:details}:
\begin{enumerate}
    \item \textbf{Number of successful runs}: The number of randomised runs that terminated "successfully" with a locally optimal solution as contrasted with the runs where the solver reported infeasibility due to being stuck in an infeasible region.
    \item \textbf{Number of distinct successful orientations}: The number of distinct orientations found from the randomised runs of the solver that terminated successfully. Two orientations are distinct if they differ in the flow direction of even a single link.
    \item \textbf{Number of common links in successful orientations}: The number of links in the network in which the flow direction is the same in each successful output.
\end{enumerate}

We observe from Table  \ref{table:details} that the 
    number of distinct successful orientations is marginally high or significantly lower and the number of common links in successful orientations is consistently higher for parallel link than discrete segment.

This suggests that the solver spends a lot less time exploring different orientations in the parallel link formulation as compared to the discrete segment formulation. Since parallel link offers a smaller standard deviation and average cost, the very few orientations it does choose to explore are clearly better. %
\fi
We can get an intuitive understanding of why Parallel Link gives a better solution 
\ifjournal
than Discrete Segment 
\fi
by observing links with near-zero flow at local optimum points. Although changing flow direction would require only a small jump in terms of value, it may require a lot bigger change in flows through many other links to compensate for the change in pressures at the two end points. We would thus like to discourage changes in flow direction through links with near-zero flow at local optimal points as it might cause the solver to wander away from an optimum when it is already near.

Parallel link achieves this through the
\ifjournal
\nameref{cons:unique_flow_dir_j}
\fi
constraint which necessitates the change in the values of both flow variables: one of them going to 0 and the other increasing from 0 to a positive value, in order to cause a change in the direction of a link.

\subsection{Re-solving parallel link}
\ifjournal
In contrast with our observations from discrete segment in Section \ref{ssec:rsds}, there is no noticeable improvement in terms of the least cost, standard deviation and average cost on re-solving parallel link. Due to this, we conclude that re-solving on parallel link is not useful and one should prefer doing more runs of the parallel link formulation from random starting points than re-solving while doing the optimization in a time constrained setting.
\fi

\section{Conclusion}

see that today's advanced solvers are increasingly allowing a shift in focus from problem specific solving techniques to a focus on refinement on formulations by employing domain knowledge of the problem. This makes today's solvers particularly useful for practical applications 
\ifjournal
such as these 
\fi
where development of novel solving techniques or even refinement of existing ones is often difficult.

\ifjournal
Secondly, the 
\fi
inability of the simple Orientation Search algorithm to give feasible orientations suggests a need to develop complex algorithmic techniques for such sub-problems. However, the success of Re-Solving also suggests that solvers are able to solve for such implicit sub-problems, as part of solving the main problem itself. This again aligns with our view that the development of highly advanced solvers has obviated the need for development of separate optimization techniques for each problem. We believe that the focus of research in this field should be on coming up with formulations and solver agnostic approaches, a philosophy we have tried to maintain throughout this paper.

\bibliographystyle{plain}
\bibliography{refs}
\end{document}